\documentclass[twocolumn,prl,10pt,amsmath,amssymb,nofootinbib,showpacs,superscriptaddress,floatfix]{revtex4-1}

\newcommand{\dfr}[2]{\frac {\displaystyle #1}{\displaystyle #2}}
\newcommand{\rs}{\rm\scriptscriptstyle}

\usepackage{graphicx}
\usepackage{color}
\usepackage[usenames,dvipsnames]{xcolor}
\usepackage[colorlinks=true,linkcolor=Red,citecolor=Green,linktoc=page]{hyperref}
\usepackage{multirow}
\usepackage{float}
\usepackage{flushend}
\usepackage{balance}
\usepackage[varg]{txfonts}
\usepackage{ulem}
\usepackage{fancyhdr}

\begin{document}
\title{Berezinskii--Kosterlitz--Thouless  and   Vogel--Fulcher--Tammann  criticality  in $\mathrm{XY}$ model}
\author{M.\,G.\,Vasin}
\affiliation{Physical-Technical Institute, Ural branch of Russian Academy of Sciences, 426000 Izhevsk, Russia}
\affiliation{Institute for High Pressure Physics of Russian Academy of Sciences, 108840 Moscow, Russia}
\author{V.\,N.\,Ryzhov}
\affiliation{Institute for High Pressure Physics of Russian Academy of Sciences, 108840 Moscow, Russia}
\author{V.\,M.\,Vinokur}
\affiliation{Materials Science Division, Argonne National Laboratory, 9700 S. Cass Ave, Argonne, IL 60439, USA}

\begin{abstract}
We develop a gauge theory of the critical behavior of the topological excitations-driven Berezinskii-Kosterlitz-Thouless (BKT) phase transition in the XY model with weak quenched disorder. We find that while in two-dimensions the liquid of topological defects exhibits the BKT critical behavior, the three-dimensional system shows more singular Vogel-Fulcher-Tamman criticality heralding its freezing into a spin glass. Our findings provide insights into the topological origin of spin glass formation.
\end{abstract}

\maketitle

\section{Introduction}
Celebrated Berezinskii-Kosterlitz-Thouless works\,\cite{Ber,KostThoul,KT1972} brought a new paradigm of topological phase transitions driven by topological excitations. At the BKT temperature, $T_{\rs BKT}$, the binding-unbinding transition between the \textit{confined} phase, at $T<T_{\rs BKT}$, where the topological excitations (vortices) of the opposite sign are bound into the `neutral' dipoles, and the unbound phase, at $T>T_{\rs BKT}$ where topological excitations unbind loose and form a `free' neutral plasma.
Among many remarkable properties of the BKT transition, the singularity of its critical behavior stands out. On approach $T_{\rs BKT}$ from above, the correlation length that sets the spatial scale for separation between the free excitation, diverges extremely rapidly, $\xi\sim\exp[\sqrt{E_0/(T-T_{\rs BKT})}]$, much faster than any power law governing the correlation length $\xi\sim|T-T_c|^{-\nu}$ for a standard continuous phase transition\,\cite{Kosterlitz1977,Jose1977}.
The BKT criticality strikingly resembles the criticality near the glass transition with the relaxation time diverging according to the Vogel--Fulcher--Tammann (VFT) law, $\tau\sim\exp[E^{\prime}_0/(T-T_{\rs G})]$. In spin glasses, the relaxation time can be related with some correlation length $\xi' \sim \tau^{1/z}$, where $z\approx 2$ is the dynamic exponent \cite{HH,PP}.
The possible connection between the two criticalities was indicated by Anderson\,\cite{Anderson1978}, who attributed the VFT criticality to logarithmic interaction between the topological excitations. However although there exists a rich lore of various specific models for both structural and spin glasses leading to VFT behavior, a universal `first-principle-like' theory for the VFT criticality is still lacking. Our paper steps into breach.

We focus on the XY model representing a wealth of physical systems ranging from Josephson junction arrays (JJA) and vortex systems in type II superconductors to spin glasses that exhibit glassy behavior. Furthermore, in two dimensions, the XY model is a generic system for the BKT transition. At the same time, the 3D XY model has been for decades an exemplary testing ground for studying a glass transition in spin systems, see, for example early papers\,\cite{DV5,DO}.
The physics of the XY-system is governed by the topological excitations (vortices). The  latter emerge in a form of vortex-antivortex pairs either due to thermal fluctuations or are induced by disorder. In the 2D system the energy to create a vortex-antivortex pair is finite thus vortices exist at any finite temperature. In the 3D systems the vortex energy is proportional to its length, therefore, only the small size vortex loops can appear as a result of thermal fluctuations, while the long-living vortex pairs of the macroscopic size in 3D may emerge only due to finite quenched disorder.
According to Dzyaloshinsky \textit{et al.}\,\cite{DV5,DO}, the stable vortices correspond to the frustration lines introduced by Toulouse and Villain\,\cite{DV8,DV7}. These frustration lines are either induced by quenched disorder\,\cite{KV}, or result from the concurrent action of system's geometrical constraints and competition between the spin interactions leading to the degeneracy of the ground state\,\cite{Nelson}.
Furthermore, quenched disorder decreases the effective dimensionality of a $d$-dimensional disordered system down to $d-2$\,\cite{SupParisi}. As a result, at certain concentration of the frustration points, one would expect that the frustrated 3D system becomes similar to a lower-dimensional system.  This implies that increasing the degree of frustration  can lead to formation of the topologically stable vortex excitations that  would exist in the low-temperature phase analogously to the vortices in the 2D system.

The glass formation in XY systems has been a subject of tireless attacks based on the quantum field theory methods\,\cite{DV5,DO,SupRivier,Hertz} and
computational approaches, see, for example,\,\cite{R44,Olsson1,Olsson2,R45}. In particular, it became possible to model the slow cooling in a 3D XY system and demonstrate the system with the initially randomly distributed spins freezes into a vortex glass. At the same time, the collinear and other anisotropic initial spin distributions evolve into a vortex-free ferromagnetic state having a lower energy \cite{R46}. Nevertheless, in spite of the substantial efforts expended, the entire vitrification process and its relation to topological excitations in XY systems still remains a mystery.
In this work we construct a gauge theory of the topological transition in the XY model. Our approach follows ideas proposed by Rivier and Dzyaloshinskii \cite{DV5,DO,SupRivier}. Unlike in the Hertz' theory\,\cite{Hertz} we consider free gauge fields induced by mobile vortices. Accordingly, the system criticality is determined by the vortex-vortex interaction.
We show that the binding-unbinding BKT transition may give rise to two distinct critical behaviors depending on the system's dimensionality.
A two-dimensional (2D) system follow the standard BKT scenario, whereas the three-dimensional (3D) system settles into a glass.

\section{$\mathrm{XY}$ model with quenched disorder in terms of the gauge field theory}
Let us consider the $d$-dimensional XY model subject to quenched disorder on a lattice, i.e. a $d$-dimensional grid with the two-component classical vector ${\bf S_r}$ of the unit length assigned to every nod ${\bf r}$. Each vector can rotate in the XY plane.
The system's Hamiltonian is:
\begin{gather}
H=-\dfr12\sum\limits_{\langle {\bf r}\neq{\bf r'}\rangle }^N\mathcal{E}_{\bf r-r'}{\bf S_r}{\bf S_{r'}}.
\label{H1}
\end{gather}
where $\mathcal{E}_{\bf r-r'}$ is the random coupling energy of nearest vectors, such that the average $\langle{\mathcal E}_{\bf r-r'}\rangle_{\bf r-r'} >0$, 
$N$ is the total number of the nodes, and brackets  $\langle {\bf r}\neq{\bf r'}\rangle$ stand for the summation over the nearest neighbors around ${\bf r}$.
The general properties of this model are well studied and depend essentially on the system dimensionality.
In $d>2$ dimensions there is a second order transition at some $T=T_c$ between the low-temperature phase and the so-called {\it symmetric phase} where spins rotate loose due to thermal fluctuations.  
In the system with the dimensionality $d\leqslant 2$, thermal fluctuations always destroy a long-range order, as
was proven by Mermin, Wagner\,\cite{Mermin} and Hohenberg\,\cite{Hohenberg} from the Bogolubov inequalities. However, the local order may hold.
Indeed, while at high temperatures the system may be fully disordered, the local ferromagnetic ordering appears upon cooling the system at some temperature $T\leqslant T_c$. This does not develop into a long-range ordering since the system contains the topological excitations.
Hereafter we will be referring to these excitations as to ``vortices'' by analogy with the vortices in superfluid helium that are also twin brothers of disclinations in the elasticity theory. In two dimensions, the Coulomb gas of
vortices experiences the BKT transition at $T=T_{\rs BKT}\equiv T_g<T_c$ the system undergoes the BKT transition. As a precursor of the transition, the correlation length, i.e. an average distance between the still unbound vortex-antivortex pair critically diverges. Yet, rigorously speaking, even at $T<T_g$ the long-range order is absent, i.e. $\langle {\bf S}\rangle=0$.

We will show below that the differences between low-dimensional and high-dimensional systems are somewhat reduced in frustrated systems.
To describe frustrated systems we develop a universal approach combining perturbative techniques applied to second order phase transition in spin system and non-perturbative methods for tackling the vortex confinement phenomenon.
We build on the Weiss theory in the long wave limit. The Hamiltonian density of the pure XY model is
\begin{gather}\label{L1}
\mathcal{H}_0=\dfr12|\nabla\Psi|^2-\dfr12m^2|\Psi|^2+\dfr b4|\Psi|^4,
\end{gather}
where the vector field $\Psi_{\bf r} =\langle {\bf S}_{\bf r} \rangle_{\delta V}=\psi_{\bf r} e^{i\Phi_{\bf r} }$, coarse grained over some finite volume $\delta V$ around the space point ${\bf r}$, plays the role of the Higgs field, $b>0$, and $m^2=\alpha (T_c-T)$ with $\alpha >0$.
Now the Hamiltonian, $H=\int \mathcal{H}_0 \mathrm{d}^dr$, where the integration is done over all volume of the $d$-dimensional system,
corresponds to the perturbative Ginzburg--Landau theory of the second order phase transition when $d>2$.

The systems described by Eq.\,(\ref{L1}) harbor topologically non-trivial equilibrium vortex excitations in the low-temperature state.
These excitations that according to the BKT theory govern the physics of the system are missed in perturbative approaches since $\nabla \Psi\sim \psi r^{-1}$ and in vortex presence the first term gives infinite contribution to the energy, $\int |\nabla\Psi|^2\mathrm{d}^dr\sim r^{d-2}\to \infty $ when $r\to \infty$.
In order to mend it, one introduces the covariant derivative instead of ordinary one, $\nabla \to D=\nabla -ig{\bf A}$, where ${\bf A}$ is the compensating field, and $g$ is the coupling constant\,\cite{Zee}:
\begin{gather}\label{L2}
\mathcal{H}_0=\dfr12|D\Psi|^2-\dfr12m^2|\Psi|^2+\dfr b4|\Psi|^4.
\end{gather}

The equilibrium value of $|\Psi|^2$ is zero at $T>T_c$ and $|\Psi|^2=\psi^2=\alpha (T_c-T)/b$ at $T<T_c$. In the low temperature region the fluctuations of the local magnetization modulus are small, $\nabla\psi \approx 0$, in addition $\nabla {\bf A}=0$. In this case one substitute $\Psi =\psi e^{i\Phi }$ in (\ref{L2}),
\begin{gather*}
\mathcal{H}_0=\dfr12\psi^2\left(\left|\nabla\Phi\right|^2-g^2{\bf A}^2\right).
\label{Lagrangean1}
\end{gather*}
If the system contains a vortex, then we can represent $\Phi$ as the sum of smooth spin waves part, $\Phi_S$, and singular vortex part, $\Phi_V$, $\Phi=\Phi_S+\Phi_V$, then ${\bf A}=g^{-1}\nabla\Phi_V$. Using the Hubbard-Stratonovich transformation and neglecting the total derivatives we can rewrite this expression in the following form:
\begin{multline*}
\mathcal{H}_1=\dfr 12{\bf B}^2
+i\psi {\bf B}\nabla(\Phi_S+\Phi_V)-\dfr12g^2\psi^2{\bf A}^2\\
=\dfr 12{\bf B}^2
+i\Phi_S\nabla(\psi {\bf B})+i\psi{\bf B}\nabla\Phi_V-\dfr12g^2\psi^2{\bf A}^2.
\end{multline*}
Integrating over $\Phi_S$, which is the Langevin source, leads to the relation $\nabla(\psi {\bf B})=\psi\nabla {\bf B}+{\bf B}(\nabla\psi)=\psi\nabla {\bf B}=0$, which is satisfied when ${\bf B}=\nabla\times {\bf a}$. The ${\bf a}$ is a gauge vector field, since this field is defined to accurate within derivative of arbitrary function $f$, ${\bf a}\to{\bf a}+\nabla f$.
It is similar to the usual vector potential of magnetic field in magnetostatics, and ${\bf B}$ is the analog of the magnetic induction. Then the Hamiltonian can be represented in the following form:
\begin{multline*}
\mathcal{H}_1=\dfr 12{\bf B}^2+i\psi(\nabla\times {\bf a})\,\nabla\Phi_V-\dfr12g^2\psi^2{\bf A}^2\\
=\dfr 12{\bf B}^2+i\nabla
(\psi({\bf a}\times \nabla \Phi_V))+i{\bf a}(\nabla\psi\times\nabla\Phi_V)\\+i\psi {\bf a}(\nabla\times\nabla\Phi_V)-\dfr12g^2\psi^2{\bf A}^2.
\end{multline*}
The second term is equal to zero if ${\bf a}\propto\nabla\Phi_V =g{\bf A}$, therefore one chooses ${\bf a}={\bf A}$. The third term is negligibly small because of the $\nabla\psi$ factor. The fourth term is non-zero only at the vortex cores where the field $\Psi_{V}$ has singularities, the gauge field source $J=J_z=2^{-1}\psi\nabla\times\nabla\Phi_V$ ($z$-axis is normal to the XY plane) and assumes discrete values since when integrating around the singularity the phase change is $2^{-1}\oint \nabla\Phi_Vd{\bf l}=J_z=\pi l$ ($l\in\mathbb{Z}$). The non-uniform equilibrium gauge field ${\bf A}$ appears only in the presence of sources. If the system contains  $n$ vortices, ${\bf J}_n$, ($n=1\dots N$), then the Hamiltonian density assumes the form:
\begin{gather}
\mathcal{H}_N=\mathcal{H}_0+i\sum\limits_{n=1}^N{\bf J}_n\cdot {\bf A},
\label{H4}
\end{gather}
where
\begin{gather*}
\mathcal{H}_0=\dfr 12(\nabla\times{\bf A})^2-\dfr 12M_0{\bf A}^2
\end{gather*}
is the free gauge field,
$M_0^2=g^2\alpha (T_c -T)/b$ is the square of gauge field mass appearing due to of Anderson--Higgs mechanism (Higgs mass).
Making use of the identity $\nabla\times(\nabla\times{\bf A})\equiv\nabla(\nabla{\bf A})-\nabla^2{\bf A}$, and the gauge symmetry, ${\bf A}\to {\bf A}+\nabla f$, one arrives at
\begin{gather*}
\mathcal{H}_0=\dfr 12{\bf A}\nabla^2{\bf A}-\dfr 12M_0{\bf A}^2.
\end{gather*}
In the momentum representation the Hamiltonian becomes
\begin{gather}
\mathcal{H}_N=-\dfr 12{\bf A}({\bf p})({\bf p}^2+M_0^2){\bf A}(-{\bf p})
+i\sum\limits_{n=1}^N{\bf J}_ne^{-i{\bf p}{\bf r}_n}\cdot {\bf A}({\bf p}),
\label{H5}
\end{gather}
and the gauge field Green function is:
\begin{gather}\label{Field}
    \langle {\bf A}({\bf p}){\bf A}(-{\bf p})\rangle_A=\dfr{-\beta^{-1}}{{\bf p}^2+M_0^2}\,,
\end{gather}
where $\langle \ldots \rangle_A$ denotes the statistical averaging over all possible $\bf A$-configurations, and $\beta=1/k_{b}T$ ($k_b$ is the Boltzmann constant).
The characteristic correlation length scale is proportional to the inverse Higgs mass, $\xi \sim M_0^{-1}$.
One immediately sees that vortices behave like currents in magnetostatics.
For example, the Hamiltonian of the system comprising two similarly charged vortices with coordinates ${\bf r}_1$ and ${\bf r}_2$, is $\mathcal{H}_{2}=\mathcal{H}_{0}+i{\bf AJ}_1-i{\bf J}_2{\bf A}$. Integrating out the field $A$, one arrives at the Biot-Savart law
\begin{gather*}
H_{2}=J^2\int \dfr{\mathrm{d}^d{\bf p}}{(2\pi)^d}\dfr{e^{i{\bf p}({\bf r}_1-{\bf r}_2)}}{{\bf p}^2+M_0^2}.
\end{gather*}
Therefore, in the 3D case
\begin{gather*}
H_{2}=\dfr{J^2}{4\pi |{\bf r}_1-{\bf r}_2|}e^{-|{\bf r}_1-{\bf r}_2|M_0}\,.
\end{gather*}
This Hamiltonian is analogous to the Hamiltonian of electric currents system in magnetostatics.

\section{Averaging over grand canonical ensemble}
In order to take account of all possible vortices configurations we utilize the Grand Canonical Ensemble description of the vortex gas. The mobile vortices screen each other, and this gives rise to the renormalization of the Higgs mass. To calculate it, one carries out the averaging over grand canonical ensemble of the ``particles'' endowed with the two possible dimensionless charges: $\sqrt{\beta a^{d-2}}J_n=\pm 1$ ($a$ is the vortex core radius).
Near the transition temperature $T_c$ the vortices are stable and mobile. As a result, the system cannot reach any ordered equilibrium state.
The partition function is:
\begin{multline*}
Z=\langle e^{ -\beta H}\rangle=\langle\langle e^{ -\beta H}\rangle_A\rangle_J=\left<\int\mathcal{D}{\bf A}
e^{ -\beta \int \mathrm{d}^d {\bf r}\mathcal{H}}\right>_J \\
=\sum\limits_{N=1}^{\infty}\dfr {\lambda^N}{N!}\sum\limits_{\{J_n\}}\int\mathcal{D}{\bf A}
\int\prod\limits_{n=1}^N \mathrm{d}^d{\bf r}_ne^{ -\beta \int \mathrm{d}^d {\bf r}\left[\mathcal{H}_0-i\delta^{(2)}({\bf r}-{\bf r}_n){\bf J}_n({\bf r}){\bf A}({\bf r})\right]},
\end{multline*}
where $\langle \ldots \rangle_J$ denotes the statistical averaging over grand canonical distribution of vortices, $\int \mathcal{D}{\bf A}$ is the functional integration over ${\bf A}$ field, $\{{\bf J}_n\}$ is the set of all configurations of ${\bf J}_n=\pm J$, $\lambda=\lambda_0e^{-\beta E_{c}}$, $\lambda_0$ is the dimensional factor of vortex density (see below), $E_c$ is the energy of the vortex core.
Note that according to the topological laws the vortex in 2D SO(2) system is the point disclination, whereas in 3D SO(2) system it is the linear one. Therefore, from the least action principle, one derives the part of the action containing the gauge field source:
\begin{gather*}
i\beta\int \mathrm{d}^d{\bf r} \delta^{(2)} ({\bf r}-{\bf r}_n){\bf J}_n({\bf r})\cdot{\bf A}({\bf r})\approx i\beta J_na^{d-2}|{\bf A}({\bf r}_n)|,
\end{gather*}
where $a$ is the vortex core radius, $J_n$ is the topological source.
Then
\begin{multline*}
Z=\sum\limits_{N=1}^{\infty}\dfr {\lambda^N}{N!}\sum\limits_{\sqrt{\beta a^{d-2}}J_n=\pm 1}\\
\int\mathcal{D}{\bf A} e^{-\beta \int\mathrm{d}^d{\bf r}\mathcal{H}_0}
\int\prod\limits_{n=1}^N \mathrm{d}{\bf r}_n\exp\left[ i\beta a^{d-2}J_n |{\bf A}({\bf r}_n)|\right].
\end{multline*}
Averaging over the dimensionless quantity $\sqrt{\beta a^{d-2}}J_n=\pm 1$ one arrives at the
\begin{multline*}
Z=\int\mathcal{D}{\bf A} e^{-\beta \int\mathrm{d}^d{\bf r}\mathcal{H}_0} \dfr {1}{N!}\sum\limits_{N=1}^{\infty}\left(2\lambda\int \mathrm{d}^d{\bf r}\cos\left[ \sqrt{\beta a^{d-2}}|{\bf A}({\bf r})|\right]\right)^N \\
=\int\mathcal{D}{\bf A} \exp\left(-\beta \int \mathrm{d}^d{\bf r} \left( \mathcal{H}_0 -2\lambda\beta^{-1}\cos\left[ \sqrt{\beta a^{d-2}}|{\bf A}({\bf r})|\right]\right)\right).
\end{multline*}
Note, that the averaging is carried out over all quantities of the point vortices and their possible positions.  This seems to be natural in the two dimensions case. However, in the 3D case only linear vortices are possible, and, at the first glance, the procedure of summing up over all configurations of all the vortex points seems incorrect.  However, while the point vortices necessarily must to be taken into account when  summing over all possible spin configurations, but are energetically unfavorable their contribution is negligible. In another words, when averaging over all the configuration of the 3D system with the proper weight, the protocol automatically ``chooses'' only linear configurations of the vortices.
Thus, the above averaging takes into account all configurations of the linear vortices including all possible loop configurations.

\begin{figure}[h]
   \centering
   \includegraphics[scale=0.48]{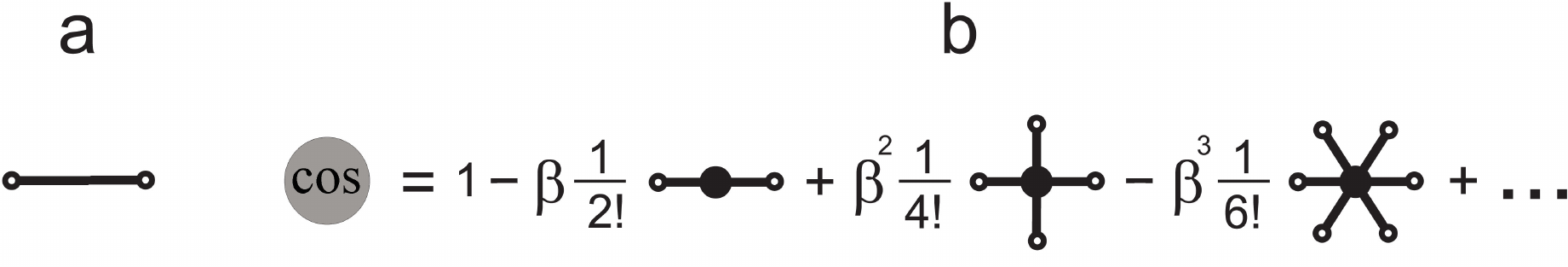}
   \caption{\textbf{Diagrammatic series.} \textbf{a:} Green function of the vector $\bf A$ field. \textbf{b:}
    Diagrammatic representation of the $\cos(\sqrt{\beta a^{d-2}}|{\bf A}|)$.}
   \label{fig1}
\end{figure}

As a result the system Hamiltonian assumes the form:
\begin{gather}
\mathcal{H}=\dfr12{\bf A}\nabla^2{\bf A}-\dfr12M_0^2{\bf A}^2-2\lambda\beta^{-1} \cos\left(\sqrt{\beta a^{d-2}}\,|{\bf A}|\right)\,,
\label{Z1}
\end{gather}
which is nothing but the Hamiltonian density of the sine-Gordon theory\,\cite{Min}. The physical meaning of $\lambda $ is the vortex density (see below). As we will see, it controls the kinetics of the system upon cooling. The graphic representation of the model's Green function and the non-linear part of the free energy density are shown in Fig.\,\ref{fig1}.

Thus, the non-linear term in Eq.\,(\ref{Z1}) renormalizes the Higgs mass, $M^2_0\to M^2={g}^2\alpha (T_c-T)/b-2\lambda a^{d-2}$ reflecting the breaking of the system's spin collinearity by the vortices at $T<T_c$. Accordingly, the critical temperature for the gauge field determined by the $ \delta^2 \mathcal{H}/ \delta A^2|_{A=0}=0$ condition, shifts from $T_c$ to $T_g=T_c-2\lambda ba^{d-2}/{g}^2\alpha $.
Eventually, one concludes that in the $T_c>T>T_g$ temperature interval the system falls into the state endowed with the local ordering, but the long-range order is still destroyed by mobile vortices. This state is referred to as the {\it disordered phase}. At $T=T_g$ the system  undergoes a phase transition, the features of which, as we show below, depend on the dimensionality of the system.
A general peculiarity of this transition is that it is the topological phase transition in which an order parameter does not arise at $T_g$ but the correlation radius of the gauge field diverges, the field becomes massive, and, as a result, at $T<T_g$ the system freezes into the state, which is named the {\it confined phase}\,\cite{Zee,Polyakov}.

We reiterate here that the above expressions are valid in the general $d$-dimensional case. However, it is well known that in pure 3D systems with degenerate continuous symmetry, the low-temperature phase practically does not contain vortices because in 3D case the energy for creating a linear vortex is indefinitely large. Thus, in the 3D system the vortices may become relevant only in the presence of the additional perturbation caused by the quenched disorder frustrating the system. The parameter characterising the degree of frustration is the vortex density, $\lambda$. Indeed, in 2D case the average vortex number at $T=T_c$ is:
\begin{multline*}
\rho = \beta \langle {\bf J}^2\rangle_{r=0} = \exp\left[-\int \dfr{\mathrm{d}^2{\bf p}}{(2\pi)^2}\dfr{1}{{\bf p}^2+M^2(T_c)}\right]\\
=\exp\left[-\ln(1+a^{-2}/2\lambda)\right]\approx a^22\lambda .
\end{multline*}
Thus, $\lambda $ is proportional to the vortex density.

If in the 3D case topological excitations are absent, then the mass of the vector field $M=M_0$ becomes zero at $T=T_c$. In this case the behavior of the XY-system does not differ from that of the Ginsburg--Landau system, and $T_c$ is the second order phase transition point. The appearance of the statistically significant number of topological excitations leads to the renormalization of the mass of the gauge field.

\section{The differences of mass renormalization in 2- and 3-dimensions}
Now we focus on the critical behavior in 2- and 3-dimensional systems. We show that it is dimensionality that defines the kinetics of the phase transition near $T_g$. To see that, let us expand the cosine in the Hamiltonian density expression in the Taylor series over the powers of $A$. The quantum field theory\,\cite{Zee} teaches us that in $2d$ case, which is marginal, all terms of the power series expansion in the system action are relevant, i.e. all orders of the perturbation theory expansion should be taken into account.
In this case the first-order approximation of the effective mass of the gauge field can be written in the following form:
\begin{gather*}
M^2_{eff}\approx M^2+2\lambda\left[ 1-\dfr 12\beta\Lambda(M^2)+\dfr 3{4!}\beta^2\Lambda^2(M^2)-\right. \\
\left. \dfr {5\cdot 3}{6!}\beta^3\Lambda^3(M^2)+\dots \right]=
M^2+2\lambda\exp \left[-\dfr {\beta}{2} \Lambda(M^2)\right],
\end{gather*}
\begin{figure}[h]
\centering
   \includegraphics[scale=0.5]{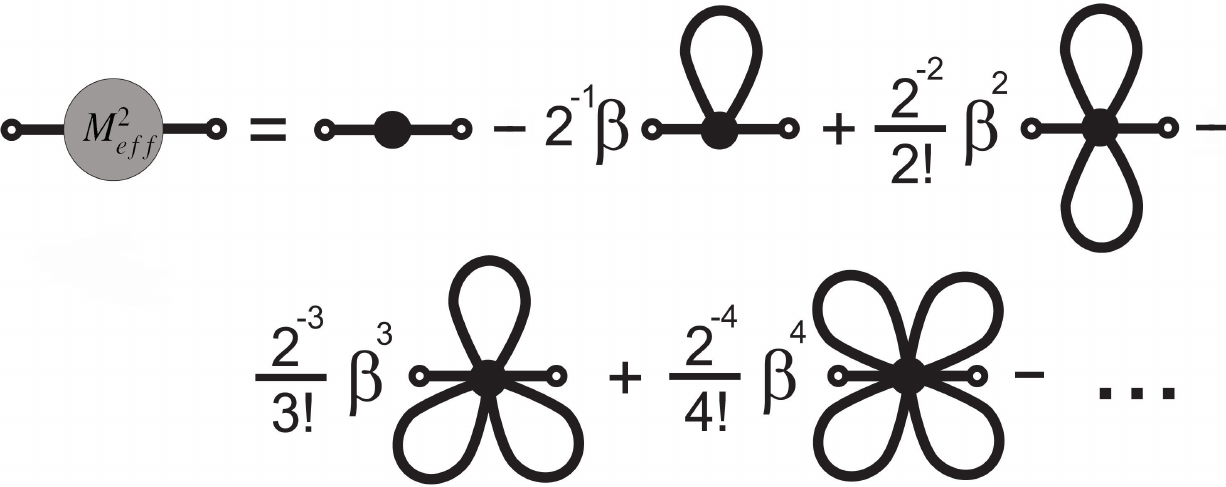}
   \caption{First-order representation of the effective mass of the vector field in the Feynman diagrams for the 2D case\,\cite{Min}.}
   \label{F1}
\end{figure}
where $\Lambda (M^2)$ is the integral corresponding to the propagator loop (Fig.\,\ref{F1}):
\begin{gather*}
\Lambda(M^2)\approx\int\limits\dfr {\mathrm{d}^2{\bf p}}{(2\pi)^2}\dfr{\beta^{-1}}{{\bf p}^2+M^2}=\beta^{-1}\ln(1+a^{-2}/M^2).
\end{gather*}
Close to the critical point $T_g$ the effective mass is small $M^2< a^{-2}$, and in the 2D case, one arrives at the effective mass given by Eq.\,(4) in the main text.
Therefore
\begin{gather}
M^2_{eff}\approx M^2+2\lambda \exp \left[-\dfr 12 \ln(a^{-2}/M^2)\right]\approx 2\lambda \sqrt{M^2a^2}.
\label{ME2}
\end{gather}

On the contrary, in three dimensions only the terms of the Taylor series expansion with powers of $A$ less than $n=6$ ($n < 2d/(d-2)$) are relevant\,\cite{Zee}. Furthermore, it is known\,\cite{Polyakov} the remaining effective nonlinearity is exponentially small, $\beta\lambda \sim e^{-\beta E_c}\ll 1$, and satisfies to the Debye approximation condition, i.e. Debye volume, $V_D$, contains sufficiently many particles in order to neglect the fluctuations of the sum of their fields.
Indeed, since the particle density $\lambda \propto e^{-\beta E_c}$, then from (\ref{Z1}) the particle number in the Debye volume, $V_D\sim(M_0^{2}+a\lambda)^{-3/2}\propto e^{3\beta E_c/2}$, exponentially diverges: $\lambda V_D\sim e^{\beta E_c/2}\gg 1$.
It means that at $d>2$ the perturbation series of the sine-Gordon theory do not contain infrared divergences\,\cite{Cvel}. Hence the system's free energy density assumes the form
\begin{gather*}
\mathcal{H}=\dfr{1}{4}{\bf A}\nabla^2{\bf A}-\dfr{M_{eff}^2}{4}{\bf A}^2,
\end{gather*}
and the effective value of the square of the vector field mass is
\begin{gather}
M_{eff}^2\approx M^2=g^2\alpha (T_{g}-T)/b.
\label{ME5}
\end{gather}

Thus, the temperature dependence of the gauge field effective mass is controlled by the system dimension and the critical behavior of 2D and 3D systems near $T_g$ appears essentially different. Note that while in both cases the correlation function $\langle {\bf A}{\bf A}\rangle$ diverges near $T_c$, the ${\bf A}$ field is not experimentally observable this divergence only on ind itself, thus this divergence manifests through the correlation functions of vortices.

\section{Correlation lengths calculation}
 The standard Landau order parameter cannot describe neither BKT transition nor the glass transition. The relevant quantity instead are the correlation functions of vortices.
In the gauge theory, at $\beta\ll 1$ and $p\to 0$ the vortex--vortex correlation function can be written as
\begin{gather*}
\langle {\bf J}({\bf p}){\bf J}({-\bf p})\rangle =\langle\langle {\bf J}({\bf p}){\bf J}({-\bf p})\rangle_J\rangle_A
 \propto \exp\left[-\beta\dfr{a^{d-4}}{{\bf p}^2+M^2_{eff}}\right].
\end{gather*}
In the disordered phase, where $M_{eff}^2<0$, the vortex correlation function decays exponentially with the distance $\mathbf{r}$,  $\langle{\bf J}(0){\bf J}({\bf r})\rangle\propto \exp(-|{\bf r}|/r_{c})$, where the correlation length, $r_c$, corresponds to the characteristic size of the vortex--anti-vortex pair. In the 3D case $r_c$ is the vortex loop size.
Then $\langle {\bf J}({\bf p}){\bf J}(-{\bf p})\rangle \propto ({\bf p}^2+r_c^{-2})^{-1}$,
and one sees that at $p\to 0$ the correlation length can be estimated in momentum representation as $r_c\propto \sqrt{\langle {\bf J}({\bf p}){\bf J}(-{\bf p})\rangle}_{p\to 0}$.
We are interested in the system behaviour at relatively hight temperatures, $T>T_g$, on the distances which appreciably exceed of the gauge field correlation length, $|M_{eff}|^{-1}<r\to\infty $. In this case at high momentums the exponential function is rapidly oscillating. Therefore, the basic contribution to the integral comes from the long wave spectrum part, $p^2\ll |M_{eff}^2|$, and
the correlation length becomes
\begin{gather}\label{RC1}
r_{c}\propto\sqrt{\langle {\bf J}({\bf p}){\bf J}({-\bf p})\rangle _{p\to 0} }\propto \exp\left[-\dfr{a^{-2}}{2M^2_{eff}}\right],
\end{gather}
which is our key general expression.

Making use of Eqs.\,(\ref{ME2}) and (\ref{RC1}), one finds the expression of the correlation length in the 2D case:
\begin{gather}
r_c\propto \exp\left[\dfr{ 1} {(2\lambda a^2)^{3/2}}\sqrt{\dfr{T_c-T_{g}}{T-T_{g}}}\right].
\end{gather}
This correlation length is the statistically averaged maximal size of the bound vortex--anti-vortex pair and reproduces the well known standard correlation radius  for the BKT transition. To find the corresponding correlation length in the 3D system we use Eqs.\,(\ref{ME5}) and (\ref{RC1}) and arrive at
\begin{gather}
r_c\propto \exp\left[\dfr{b }{2\alpha (ag)^2}\dfr{1}{T-T_{g}}\right]=\exp\left[\dfr{1}{2\lambda a^{3}}\dfr{T_c-T_{g}}{T-T_{g}}\right].
\end{gather}
This divergence is more singular than the BKT criticality and is exactly the VFT behavior, characteristic to the glass transition.
Note that usually the critical behavior of the glasses is described in terms of the relaxation time. Here in the critical dynamics of the XY-model, the correlation length growth is accompanied by the corresponding divergence of the relaxation time, $\tau \sim {r_c}^z$ \cite{HH,PP}, which implies the critical slowing down of both 2D and 3D systems near $T_g$. We remind that in both cases the long-range correlation of vortices does not mean any long-range ordering of spins themselves, i.e. $\langle \Psi\Psi\rangle|_{r\to\infty} = 0$. This establishes that in frustrated 3D system there is a topological phase transition analogous to the BKT transition in 2D system. As we show below, this is the glass transition.

\section{Evidence of glassiness}

 Since our consideration is restricted to the static theory, we cannot conclusively judge on the system ergodicity below $T_g$ in order to prove the system's glassiness. Neither our theory holds in the non-ergodic glass state.  Thus, we have to seek some additional arguments to support our conclusion that at the temperature $T_g$ the system freezes into a glass.

First of all we note that because of the presence of stable vortices, any long-range order in the low-temperature phase is destroyed, $\langle \Psi \rangle =0$. The spin correlation function $\langle\Psi\Psi\rangle|_{r\to\infty}=0$, and the spin correlation length is small, $\sim m^{-1}=\sqrt{g/2\lambda b a^{d-1}}$.
As we mentioned above, by making use of the dispersion relation, $\omega \propto p^z$, for  the relaxation dynamics, one can arrive at the relaxation time described by the VFT dependence: $\tau \propto {r_c}^z\propto\exp \left[E^{\prime}_0(T-T_{g})^{-1}\right]$.  Hence one can suppose that the transition at $T=T_g$ belongs in the same universality class as the glass transition in the elastic media\,\cite{Nelson}. These properties are the characteristic features of the glass transition.

\begin{figure}[h]
\centering
   \includegraphics[scale=0.4]{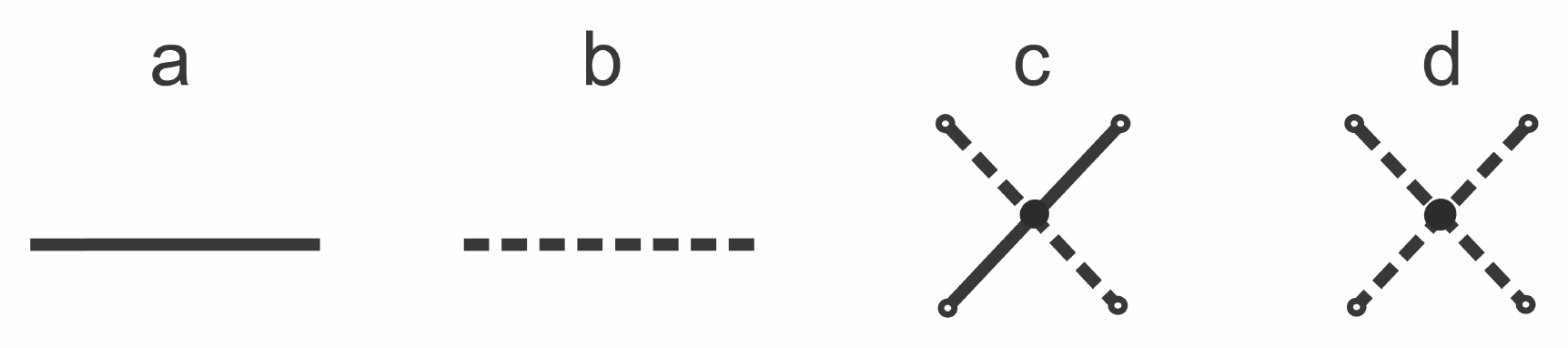}
   \caption{a) is the Green function of the vector field $\bf A$ field,
    b) is the Green function of the order parameter $\Psi $, c) is the diagrammatic representation of $g^2$, and d) is the diagrammatic representation of $b$.}
   \label{Fig 1}
\end{figure}

Another criterion of glass transition in spin system is the characteristic behavior of the linear and non-linear  susceptibility at the transition temperature \cite{binder}.
In order to determine these properties we add to the Hamiltonian of our system some external field ${\bf h}$:
\begin{gather*}
\mathcal{H}= \dfr12{\bf A}\nabla^2{\bf A}+\dfr12|\nabla \Psi|^2-\dfr12{g}^2{\bf A}^2|\Psi|^2-\dfr12m^2|\Psi|^2+\dfr b4|\Psi|^4-{\bf h}\Psi.
\end{gather*}
The Green functions of the model's fields are (\ref{Field}), and
\begin{gather*}
\langle\Psi({\bf p})\Psi(-{\bf p})\rangle =\dfr{\beta^{-1}}{{\bf p}^2-m^2}.
\end{gather*}
The Green functions and vertices of the model may be graphically represented in the form shown in Fig.\,\ref{Fig 1}.

\begin{figure}[h]
\centering
   \includegraphics[scale=0.3]{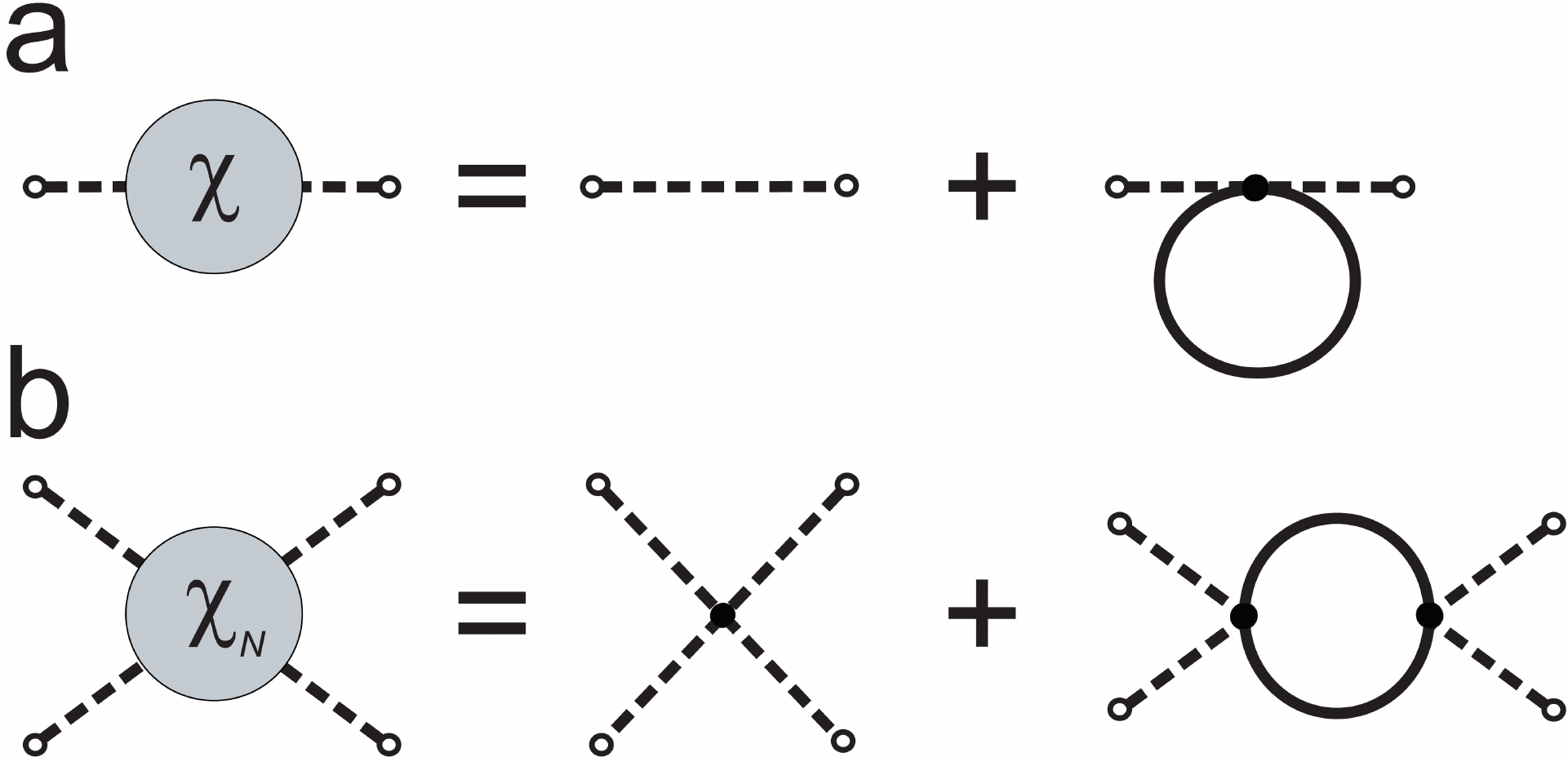}
   \caption{a) The graphical presentation of the one-loop approximation of the linear susceptibility. One can see that near the $T_g$ this value depends generally on the renormalization of $m^2$, that does not lead to appearing of any divergences. b) The graphical presentation of the one-loop approximation of the nonlinear susceptibility.}
   \label{S}
\end{figure}

The linear susceptibility of the system is $\chi_L =\left.\partial \langle\Psi\rangle/\partial {\bf h}\right|_{h\to 0,\,p\to 0}=\beta\left.\langle\Psi^2\rangle\right|_{p\to 0}$ ($h$ is an external source of the field $\Psi $).
Close to $T_g$ this value is renormalised. In one-loop approximation it can be represented in the diagrammatic form shown in Fig.\,\ref{S} a.
Using the $d$-dimensional integration property~\cite{Coll}:
\begin{gather*}
   \int \mathrm{d}^d{\bf p}\dfr{({\bf p}^2)^{\alpha}}{({\bf p}^2+M^2)^{\gamma}}=\pi^{d/2}M^{d+2\alpha-2\gamma}\dfr{\Gamma(\alpha+d/2)\Gamma(\gamma -\alpha-d/2)}{\Gamma(d/2)\Gamma(\gamma)},
\end{gather*}
one can conclude that the Higgs field loop gives the following contribution:
\begin{gather}\label{pere}
   \approx g^2\beta \int  \dfr{\mathrm{d}^3{\bf p}}{(2\pi)^3}\dfr{\beta^{-1}}{{\bf p}^2+M^2(T)}=  -g^2|M(T)|.
\end{gather}
Therefore the second term in this expression is small since it is proportional to $M$:
\begin{gather*}
\chi_L=\beta\langle\Psi^2\rangle_{p=0}=\beta\langle\Psi^2\rangle_{p=0}^0+ \beta|M|g^2\left[\langle\Psi^2\rangle_{p= 0}^0\right]^2+\dots \,.
\end{gather*}
As a result $\chi_L\approx \beta^{-1}m^{-2}=[\alpha (T_c-T_g)/T_g]^{-1}$. This value is finite in $T_g$, that satisfies to the glass transition, unlike the infinitely divergent value at the second order phase transition.

The nonlinear susceptibility is $\chi_N=\left.\partial^3\langle \Psi\rangle/\partial {\bf h}^3\right|_{h\to 0,\,p\to 0}=\beta^3\langle \Psi^4\rangle
_{p=0}$.
In the one loop approximation the renormalization of this value can diagrammatically be represented in the  form shown in Fig.\,\ref{S}~b. Close to $T_g$ the second term gives diverging contribution which dominates. As a result the nonlinear susceptibility can be estimated as $\chi_N\propto -\ln(T-T_g)$ for $T\to T^{+}_g$.
This value diverges at $T=T_g$, that also corresponds to the glass transitions in the spin systems~\cite{binder}.

The combination of the finite linear susceptibility with the infinite non-linear susceptibility in $T_g$ coupled with the destroyed long-range order, $\langle \Psi \rangle =0$, below $T_g$ specifies that $T_g$ is the glass transition temperature, see \,\cite{binder,Kivelson,K2,SupRivier,R2,Nussinov,Vasin}.
The considered physical picture also agrees with the ‘frustration-limited domain theory~\cite{Kivelson,K2}, and with `gauge theory of glass~\cite{SupRivier,R2}.
Besides, earlier, using the non-equilibrium critical dynamics methods, we have shown that the frustrated 3D system, which undergoing the second order phase transition or weak first order phase transition, does not reach the low-temperature ordered state, but freezes  in the non-ergodic glass state~\cite{SupVasin}. Therefore, we can conclude that $T_g$ is indeed the glass transition temperature.

\section{Frustration effects on the critical temperature of the second order phase transition in 3D model}


Finally, let us consider a 3D weakly frustrated system with the low vortex density, so that the frustration would not eliminate the second order phase transition.
At $T=T_c$  the nonzero local magnetization arises, $|\Psi|^2=\psi^2$, but the frustration generates vortices that destroy the order on large scales. As a result, in the some temperature interval below $T_c$ the phase transition still does not occur, since in this temperature interval the correlation radius, $r_c$, of the correlation function $\langle \Psi\Psi\rangle_{\bf r}$ remains finite.
If the vortex density is low, then the phase transition occurs nevertheless, but at temperature $T<T_c$.
In other words, the low density vortices renormalize the phase transition temperature, and shift it downwards, $T_c\to T_c^R$. Note, that in contrast to the glass transition, here the diverging quantity is the order parameter correlation length.
\begin{figure}[h]
\centering
   \includegraphics[scale=0.4]{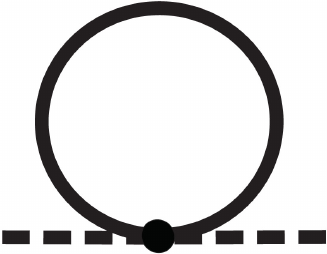}
   \caption{The one-loop contribution in the Higgs field mass renormalization.}
   \label{MR}
\end{figure}
The diagrammatic representation of the one-loop contribution in this renormalization is given in Fig.\,\ref{MR}. As well as in the above susceptibility expression\,(\ref{pere}) the Higgs field loop gives $\approx  \beta^{-1} g^2|M(T)|$.
Therefore
\begin{multline*}
   \alpha(T-T_c^R)\approx \alpha(T-T_c)+\beta^{-1}g^2|M(T_c)|\\
   = \alpha(T-T_c)+\beta^{-1}g^2\sqrt{g^2\alpha (T_c-T_{g})/b}.
\end{multline*}
Since $M^2(T_g)=g^2\alpha (T_g-T_c)/b+2\lambda a^{d-2}=0$, the new phase transition temperature depends on the vortex concentration, $\lambda $:
\begin{gather}\label{RPT}
T_c^R(\lambda )=T_c-\dfr{g^2}{\beta\alpha}\sqrt{2a\lambda}.
\end{gather}
 If at some vortex density $\lambda $ the inequality $T_c^R(\lambda)>T_g(\lambda)$ holds, then the system experiences the second order phase transition, at $T=T_c^R$. In the opposite case,  $T_c^R(\lambda)<T_g(\lambda)$, the system undergoes the transition to confined phase (glass transition) at $T=T_g$.
The sketch of the resulting phase diagram is shown in Fig.\,\ref{PD3}.
\begin{figure}[h]
\centering
   \includegraphics[scale=1.2]{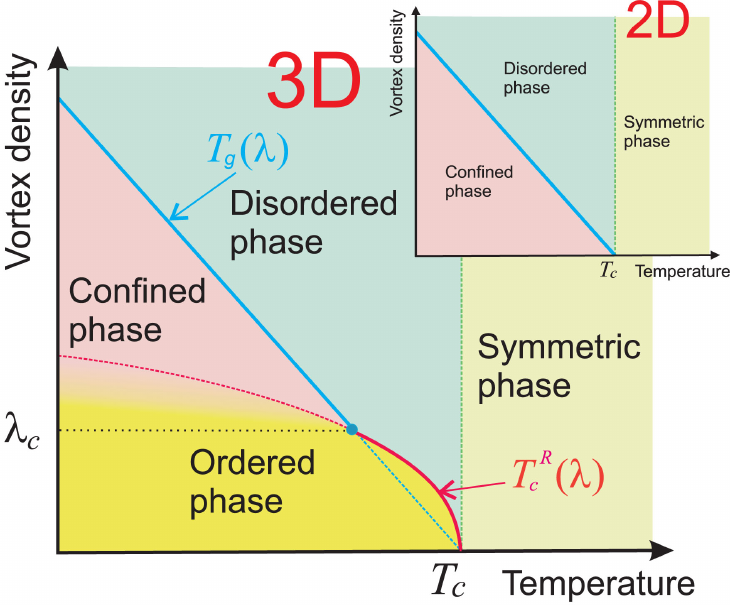}
   \caption{\textbf{Schematic phase diagrams of the vortex systems in the frustrated 3D and 2D $\mathrm{XY}$-models in the $T$--$\lambda$ coordinates.} The red line marks the renormalized temperature of the second order phase transition, $T_c^R(\lambda )$, and blue line, $T_g(\lambda )$, is the dependence of the temperature of the confinement transition of $\lambda $.  At $T_c^R>T_g$ the 3D system undergoes the second order phase transition. At  $T_c^R<T_g$ the system undergoes the transition to the confined phase. The dashed red and blue lines are prolongations of the red and blue solid lines into the zone of the solid phases. }
   \label{PD3}
\end{figure}

\section{Discussion}

We constructed a gauge quantum theory of the XY-model with quenched disorder taking into account formation of vortex-like toplogical excitations. We have found that at $T=T_c$ the disordered XY-model acquires an orientational stiffness\,\cite{PP}, which leads to the formation of vortices. We extended the BKT approach onto the systems subject to quenched disorder and demonstrated that in two dimensions the XY-model experiences the customary BKT transition at $T=T_g$ into a confined phase, whereas the three-dimensional XY-model, where vortices appear in a form of the vortex lines, i.e. vortices starting and ending at the system surfaces, and vortex loops of an arbitrary size\,\cite{Obukhov}, undergoes the glass transition that exhibits the Vogel-Fulcher-Tamman critical behavior. However, if the vortex concentration is not high enough, and remains below some threshold value, the system undergoes the usual second order phase transition.


The results are summarized in the phase diagram in Fig.\,\ref{PD3}.
The second order transition is shown by the solid red line.
At relatively strong quenched disorder and, as a result, high enough vortex density, the system falls into the disordered confined phase, where the movement of the vortex is limited by the frustration and by other vortices (e.g. vortices get entangled).  Furthermore, the increase in dimensionality reduces the effective strength of thermal fluctuations so that they may not be able to ``push apart'' the entangled vortices and drive the frustrated system into the ordered state.  As a result, the strongly frustrated 3D system with the vortex density exceeding some critical value $\lambda_c$, freezes into the disordered confined phase at $T_g(\lambda )>T_c^R(\lambda )$. The transition is shown as the blue solid line in Fig.\,\ref{PD3}.

\begin{figure}[h!]
   \centering
   \includegraphics[scale=0.35]{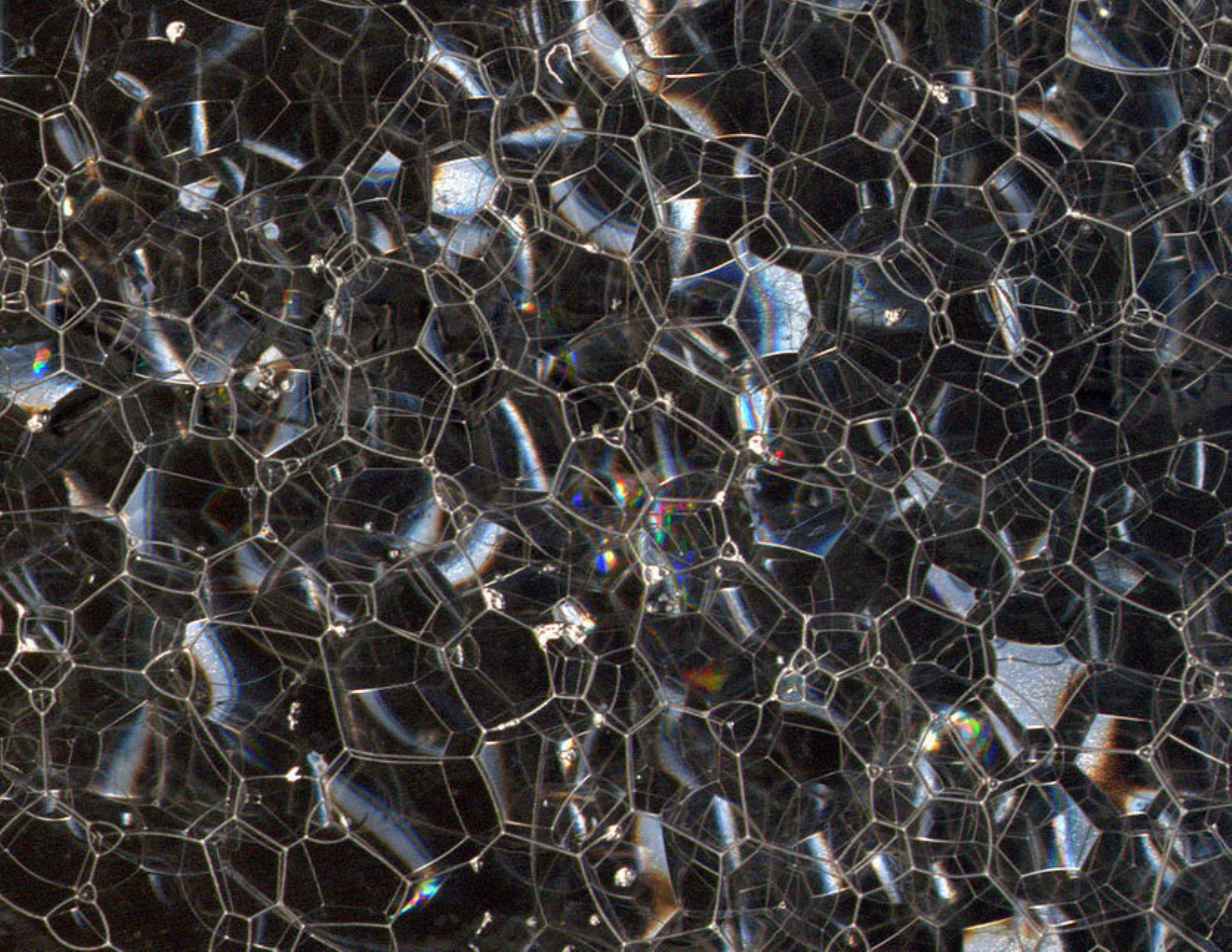}
   \caption{\textbf{The soap suds illustrating the configuration of the gauge foam formed by the interacting linear vortex network in the confined phase in 3D}. The edges in the foam correspond to the vortex lines in the confined phase. In these lines the gauge field and the spin vector field have singularity.  The vortex net energy is proportional to the total area of membranes connecting the vortex lines, like the energy of the soap suds is proportional to the total area of the soap film.}
  \label{foam}
\end{figure}

Both 2D and 3D systems undergo a phase transition from disordered phase to the confined phase at $T=T_g$, where the gauge field ${\bf A}$ becomes massive so that $(\Delta+M^2){\bf A}=0$. If $M^2<0$, this field is screened at the distance $|M|^{-1}$, as an analogue to the Meisner effect.  At the same time, $\nabla \cdot{\bf A}=0$ regardless the finiteness of the vector field mass. To reconcile both conditions, the field has to form strings in the 2D case, or membranes in the 3D case, which confine the field flowing from the `source' charge into the `drain" one of the opposite sign. In quantum field theory this phenomenon is known as {\it confinement}~\cite{Polyakov}.
In the confined phase the energy of the gauge field is concentrated within the field membranes, thus the field energy is proportional to the membrane's total area, and the confined phase can be viewed as a gauge ``foam'' (see Fig.\,\ref{foam}). One can conclude that in 3D this phase is a spin glass, since it is disordered and has the glass state attributes.

\section{Conclusions}

One of the main results of this work is that in XY-model with the quenched disorder, both BKT transition in 2D and the glass transition in 3D systems are of the same topological nature.
The required condition for the topological transition in 3D system is the presence of quenched disorder reinforcing the thermal fluctuations.
The above consideration was formally restricted to the glass transition in the framework of the XY-model, i.e. to a subclass of spin glasses, our conclusions apply also to more broad class of the glass-forming systems. At the same time there are challenges to address, especially concerning the behavior of the structural glasses. In the latter, the glass transition is significantly more complex, since disorder can be self-induced during the system's freezing because of geometrical frustrations~\cite{Nelson,R2}. However, this also points out to the complex topological nature of this process, and its ascertainment is one of the important directions of future investigations.

Note that the topological phase transition associated with the appearance of topologically stable linear perturbations (topological defects) of infinite length is well known as the crystal melting model \cite{R42}. We have shown that, inherently, the transition to the glass state from the high-temperature disordered phase is also a topological phase transition. There is, however, an essential difference between the topological melting and the glass transition. In the former case, the special topologically stable perturbations arise as a result of thermal excitation, while in spin glasses they also exist at low temperatures, due to quenched disorder, in a state with a frozen configuration \cite{DV5}.

Our results are of a general character with the far-reaching implications going well beyond the immediate context where they were derived.
It is noteworthy a remarkable resemblance of the evolution of the frustrated XY-model governed by formation of vortices and the Kibble-Z\"{u}rek scenario of the universe evolution. The essential feature of the latter is also the creation of topological defects due to colliding rapidly growing nuclei of the `cold' phase endowed with the different orientations. One can conjecture that in the 3D universe, the entanglement effects would lead to the exponentially long times necessary to anneal the resulting glass state.
Carrying  this analogy further, one might expect, in turn, that in the course of the universe cooling during its expansion, its continuous symmetry can break down  leading to formation of cosmological strings.
The latter thus is the result of an initial frustration of the system because the phase of the frozen-out field configuration is determined independently in the regions which did not have a chance to interact\,\cite{Kibble,Zurek}.
The network of these strings that evolves under the combined effects of tension and interaction with matter, and the velocity of the matter circulation is proportional to the inverse square of the distance between (``domain'' size). Since the universe is (3+1)-dimensional, one would expect that the system of the strings, i.e. the universe falls into a glass state where, in particular, one does not have a notion of the universal ``universe's temperature''. One might rather describe the temperature in terms of the inhomogeneous field of the local temperatures in the specific domains. This would imply that the quasi-equilibrium Boltzmann statistics holds only on the relatively small scales, but on the large time and space scales the statistic substantially changes, the system kinetics becomes a glass-like one and is characterized by the exponentially broad spectrum of relaxation times.

Note, finally, that, in prospect, the constructed gauge theory, combined with the critical dynamics methods, will enable the study of the nonequilibrium dynamics near the topological phase transition both in 3D and 2D cases and the description of the dynamically-induced heterogeneity during glass formation.

\subsection{Acknowledgments}
We are grateful to Vikram Tripathi for
stimulating discussions. The work was supported by the U.S. Department of Energy,
Office of Science, Materials Sciences and Engineering Division (VV) and by Russian Foundation for Basic Research, Grants 17-02-00320 (VNR) and 18-02-00643 (MV).



\end{document}